%% file: ephs-mtns2022-article.tex
\documentclass{ifacconf}

\usepackage{graphicx}
\graphicspath{{figures/}}

\usepackage{natbib}

\usepackage{amsmath}
\usepackage{amssymb}
\usepackage{physics}
\usepackage{quiver}
\usepackage{adjustbox}

\begin{document}

\begin{frontmatter}
	\title{EPHS: A Port-Hamiltonian Modelling Language}

	\author[FAU]{Markus Lohmayer}
	\author[FAU]{Sigrid Leyendecker}

	\address[FAU]{%
		Friedrich-Alexander University Erlangen-Nürnberg,
		Erlangen, Germany
		(e-mail: markus.lohmayer@fau.de).
	}

	\begin{abstract} 
		\input{abstract.tex}
	\end{abstract}

\end{frontmatter}

\section{Introduction}%

Mathematical modelling of physical systems
and the closely related fields of
computer simulation,
optimization,
and control
keep growing in their importance for
scientific discovery
and the design and operation of engineered systems.
%
Although compositional thinking is prevalent
throughout science and engineering,
modelling and simulation software
is still lacking in this respect.
Contrary to reality,
models are often restricted to isolated systems
and model specifications are often intermingled with
the computational procedures used to evaluate them.

The advent of
equation-based modelling languages like Modelica
has improved the situation
by offering a declarative approach:
Components are defined by their
time-dependent variables,
parameters,
connectors,
and the equations relating the former.
Hierarchical models can be defined by
instantiating components and connecting them via their connectors.
The compiler turns model specifications into
(hybrid) systems of differential-algebraic equations (DAEs)
and uses structural simplification algorithms
to make them amenable to numerical methods for simulation,
see~\cite{2021Modelica}.

Some limitations
however prompt us to think that
future tools for modelling physical systems
might have to rely on a different kind of language:
While Modelica and similar languages offer
a practical way to write down systems of DAEs,
they lack the ability to
express models with more abstract semantics
such as those defined via
partial differential(-algebraic) equations
stated within the vector or exterior calculus formalism.
Further, these languages are unable to
directly express and assert
the structural properties inherent to
all models of physical systems.
%
This structure can be made manifest via
the Port-Hamiltonian Systems (PHS) framework,
see Sec.~\ref{ssec:phs}.
%
Despite of successful efforts to
integrate the PHS framework with Modelica,
see~\cite{2020MarquezZufiriaYebra},
we want to formalize a new language
specifically tailored towards
expressing (lumped- and distributed-parameter) port-Hamiltonian semantics
in a simple and mathematically rigorous manner.
We seek a language which conveys
the physical and compositional structure
at a level above the defining equations,
where it can be exploited more easily
for the purpose of
analysis, model transformations, simulation, optimization, control,
and scientific machine learning.

The present paper is concerned with
the formalization of a language for
composable networks of multiphysical and thermodynamic systems
within the Exergetic Port-Hamiltonian Systems (EPHS) framework,
see~\cite{2021LohmayerKotyczkaLeyendecker}.
%
Due to its intuitive diagrammatic syntax,
the proposed language could form
a basis for novel computer-aided engineering (CAE) tools
as well as facilitate
interdisciplinary communication, decision making, and education.
A close connection to the exergy analysis method
makes the language particularly interesting for
thermodynamic optimization and sustainable engineering.

Throughout science and engineering,
various sorts of diagrams,
such as circuit diagrams, block diagrams, and Petri nets,
are employed to
depict information about mathematical models.
In applied category theory research,
the formalization of such diagrams
based on monoidal categories and their underlying operads
has been a major topic,
see e.g.~\cite{%
2015BaezErberle,%
2017BaezPollard,%
2018BaezFong,%
2021BaezCourserVasilakopoulou,%
2021BakirtzisVasilakopoulouFleming%
}
and references therein.
We focus on the operadic perspective
because it is much closer to
how we visually express and implement the language.
An operad is a mathematical object
used to organize formal `operations'
which have finitely many `inputs' and one `output'.
An operad algebra then endows
the formal operations with semantics.
One speaks of a coloured operad
when its operations compose only if
their respective inputs and outputs have matching types,
see~\cite{2004Leinster,2016Yau}.
We shall make use of
the operad of undirected wiring diagrams
which initially appeared in~\cite{2013Spivak}
together with its relational algebra
used to model database queries,
see also~\cite{2014Spivak,2018Yau,2020BreinerPollardSubrahmanianMarie}.
The recent paper by~\cite{2021LibkindBaasPattersonFairbanks}
deals with an operadic approach to
modelling of discrete and continuous dynamical systems
with directed and undirected semantics of composition
and its software implementation.

Relying on the operad of undirected wiring diagrams,
we can give a precise definition of
a diagrammatic syntax
for (exergetic) port-Hamiltonian systems
which has a close connection to bond graphs.
Further, we can formalize different kinds of
(exergetic) port-Hamiltonian semantics
as an algebra over this operad.

To make the paper as self-contained as possible,
in Sec.~\ref{sec:background} we review some concepts
from port-Hamiltonian systems and category theory.
In Sec.~\ref{sec:language} we introduce
the syntax of the proposed modelling language
from an intuitive standpoint
and then go on to show how it can be formalized
based on the operad of undirected wiring diagrams.
Finally we briefly mention how to assign semantics.
In Sec.~\ref{sec:conclusion}
we state our conclusions
and give an outlook on future work.

\section{Background}%
\label{sec:background}

We introduce mostly via example
necessary concepts from
(exergetic) port-Hamiltonian systems
and category theory.

\subsection{Port-Hamiltonian Systems}%
\label{ssec:phs}

For didactic purposes,
we consider one of the simplest examples:
A damped harmonic oscillator
features a spring, a moving mass and a damper.
The stored energy is expressed via
the Hamiltonian function $H$ defined by
\begin{equation}
	H(q, p)
	\: = \:
	E_\text{spring}(q) + E_\text{mass}(p)
	\: = \:
	\frac{1}{2 \, c} \, q^2 + \frac{1}{2 \, m} \, p^2
	\,.
	\label{eq:osc_H}
\end{equation}
The extension $q$ and the momentum $p$ are state variables,
while the compliance $c$ and the mass $m$ are fixed parameters.
The centrepiece of a port-Hamiltonian system (PHS)
is its Dirac structure
which can be defined via a skew-symmetric matrix
encoding the power-conserving interconnection:
\begin{equation}
	\left[
		\begin{array}{c}
			\dot{q} \\
			\dot{p} \\
			\hline
			e_{Rm}
		\end{array}
	\right]
	\: = \:
	\left[
		\begin{array}{rr|r}
			0 & 1 & 0 \\
			-1 & 0 & -1 \\
			\hline
			0 & 1 & 0
		\end{array}
	\right]
	\,
	\left[
		\begin{array}{c}
			\partial_q H \\
			\partial_p H \\
			\hline
			f_{Rm}
		\end{array}
	\right]
	\label{eq:osc_Dm}
\end{equation}
The damping is expressed through
a resistive relation $\mathcal{R}$
which relates the flow and effort variables
of the third port:
\begin{equation}
	f_{Rm}
	\: = \:
	d \, e_{Rm}
	\label{eq:osc_Rm}
\end{equation}
The Dirac structure implies the power-balance equation
\begin{equation}
	\partial_t \bigl( H(q(t), p(t)) \bigr)
	\: + \:
	e_{Rm}(t) \, f_{Rm}(t)
	\: = \: 0
	\label{eq:osc_pbe}
\end{equation}
which says that
the net discharge of the stored energy equals
the power which is `dissipated' in the damper.
According to our sign convention,
the stored power,
the dissipated power,
and the power supplied via additional boundary ports
sum to zero.
If the Hamiltonian is bounded from below,
the system is said to be passive.
The ways how
components of a model
can exchange power
is often expressed schematically as a bond graph:
\[\begin{tikzcd}
	{H_\text{spring} : \mathbb{C}} & {\mathcal{D}} & {H_\text{mass} : \mathbb{C}} \\
	& {\mathcal{R} : \mathbb{R}}
	\arrow["{\dot{q}}", "{q/c}"', harpoon, from=1-2, to=1-1]
	\arrow["{\dot{p}}"', "{p/m}", harpoon', from=1-2, to=1-3]
	\arrow["{f_{Rm}}", "{e_{Rm}}"', harpoon, from=1-2, to=2-2]
\end{tikzcd}\]
$\mathcal{D}$ represents the Dirac structure,
$\mathbb{C}$ (like capacitor) symbolizes storage,
and $\mathbb{R}$ (like resistor) stands for dissipation.

We refer to~\cite{2014SchaftJeltsema} for
a more in-depth introduction to port-Hamiltonian systems.

\subsection{Exergetic Port-Hamiltonian Systems}%
\label{ssec:ephs}

The EPHS framework establishes
a more rigorous thermodynamic foundation for PHS:\@
Rather than energy, the central notion is exergy,
a quantity which indeed is dissipated in thermodynamic systems.
Its definition relies on a (reference) environment
which serves to quantify the degradation of energy.
It further acts as an isothermal reservoir which can absorb waste heat
and as an isobaric atmosphere, etc.
The word `environment' thus acquires a meaning
quite different from `other systems'.

In place of~\eqref{eq:osc_Rm},
the resistive structure is defined by
\begin{equation}
	\left[
		\begin{array}{c}
			f_{Rm} \\
			f_{Rt}
		\end{array}
	\right]
	\: = \:
	\frac{1}{\theta_0} \, d \,
	\left[
		\begin{array}{rr}
			\theta_0 & -\upsilon \\
			-\upsilon & \upsilon^2 / \theta_0
		\end{array}
	\right]
	\,
	\left[
		\begin{array}{c}
			e_{Rm} \\
			e_{Rt}
		\end{array}
	\right]
	\,,
	\label{eq:osc_R}
\end{equation}
where
$\upsilon := p / m = e_{Rm}$ is the velocity of the mass
and $\theta_0$ is the fixed reference temperature of the environment.
It relates the mechanical port
with an additional thermal port
in an energy-conserving,
yet dissipative, i.e.~entropy producing, manner.
The symmetry of the above matrix
is related to Onsager's reciprocal relations.
Its positive semidefiniteness encodes
non-negative entropy production
and its non-trivial kernel encodes
conservation of energy.
To express the disposal of waste heat,
the thermal port is connected to the environment
via an extension of the Dirac structure:
\begin{equation}
	\left[
		\begin{array}{c}
			\dot{s}_e \\
			\hline
			e_{Rt}
		\end{array}
	\right]
	\: = \:
	\left[
		\begin{array}{r|r}
			0 & -1 \\
			\hline
			1 & 0
		\end{array}
	\right]
	\,
	\left[
		\begin{array}{c}
			0 \\
			\hline
			f_{Rt}
		\end{array}
	\right]
	\label{eq:osc_Dt}
\end{equation}
Here,
$s_e$ is the entropy state variable of the environment.
The effort variable corresponding to $\dot{s}_e$ is $0$
because heat exchanged with the environment has no exergetic value.
Eqs.~\eqref{eq:osc_H},~\eqref{eq:osc_Dm},~\eqref{eq:osc_R},~\eqref{eq:osc_Dt}
defining the EPHS
reduce to
$\dot{q} = \upsilon$,
$\dot{p} = -\frac{1}{c} \, q - d \, \upsilon$, and
$\dot{s}_e = \frac{1}{\theta_0} \, d \, \upsilon^2$.
Even though this is more concise,
it is less amenable to analysis.
The structured representation brings to light
the inherent properties of thermodynamic systems,
first and foremost
conservation of energy and non-negative entropy production.
Passivity is tantamount to
stability of the thermodynamic equilibrium state.
We refer to~\cite{2021LohmayerKotyczkaLeyendecker} for more details.

\subsection{Category Theory}%
\label{ssec:ct}

A category has
objects
and (homo)morphisms
also called arrows. %
We write $\mathrm{Ob}(C)$
for the set of objects of the (small) category $C$
and for any $x, \, y \in \mathrm{Ob}(C)$
we write $\mathrm{Hom}(x, \, y)$
for the set of morphisms
with domain object $x$
and codomain object $y$,
i.e.~for arrows from $x$ to $y$.
Morphisms compose associatively.
For $f \in \mathrm{Hom}(x, \, y)$
and $g \in \mathrm{Hom}(y, \, z)$
there exists
$g \circ f \in \mathrm{Hom}(x, \, y)$.
For every
$x \in \mathrm{Ob}(C)$,
there exists
a designated morphism
$\mathrm{id}_x \in \mathrm{Hom}(x,x)$
which behaves as an identity
with respect to (pre- and post-)composition.
We write this as the diagram
\[\begin{tikzcd}
	x & y & z
	\arrow["f", from=1-1, to=1-2]
	\arrow["g", from=1-2, to=1-3]
\end{tikzcd}.\]
Composite and identity morphisms
are usually omitted because
their existence is guaranteed by the axioms.
Every directed graph can be made into a category
with vertices as objects
and edges as morphisms.
Identity and composite morphisms
must be (freely) added.
String diagrams are (Poincaré) dual
to diagrams such as the above
and provide a formal syntax for morphisms in (monoidal) categories:
\begin{center}
	\includegraphics[width=4.5cm]{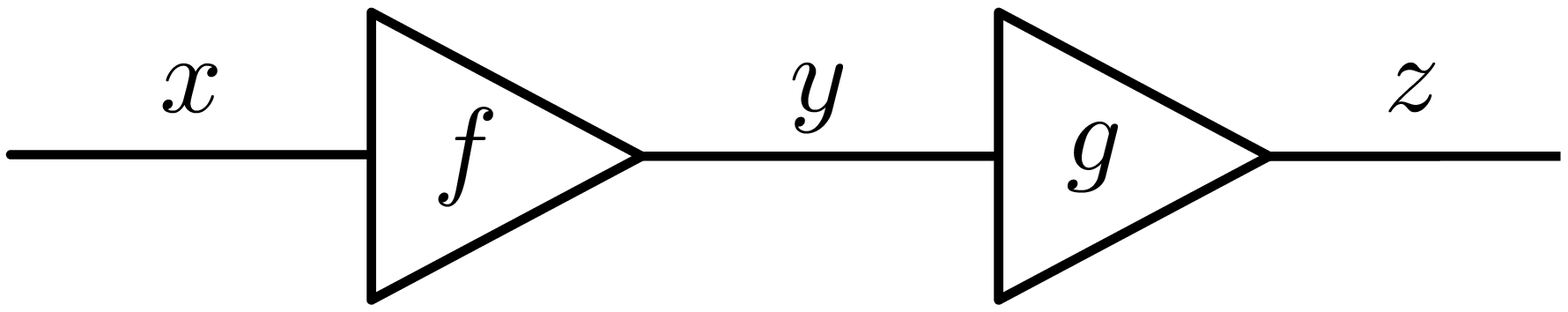}
\end{center}
Objects and their identity morphisms
are drawn as strings (edges),
and the morphisms are drawn as boxes (nodes).
An import (small) category is $\mathrm{FinSet}$
whose objects are finite sets
and whose morphisms are functions between them.
There is also a category $\mathrm{Set}$ of
(`small' but not necessarily finite) sets
and functions between them.
(This category is not small because
its objects do not form a set.)
%
A functor is a structure-preserving map between categories.
If $C$ and $D$ are categories,
a functor $F: C \rightarrow D$
maps an object $x \in \mathrm{Ob}(C)$
to an object $F(x) \in \mathrm{Ob}(D)$
and maps a morphism $f \in \mathrm{Hom}_C(x, \, y)$
to a morphism $F(f) \in \mathrm{Hom}_D(F(x), \, F(y))$.
A construction is functorial if it can
be expressed as a map between categories
which preserves composition and identity.
A prominent example is the tangent functor
which maps the category of smooth manifolds
to the category of smooth vector bundles.
It sends a manifold to its tangent bundle
and it sends a map between manifolds to its linearisation.
Functoriality is tantamount to the chain rule,
i.e.~the linearisation of a composite
is equal to the composite of the linearisations.

Multicategories generalize categories by allowing
morphisms with finitely many domain objects.
A multicategory is called symmetric if
permuting the order of the domain objects
plays together nicely with composition.
Symmetric multicategories are also called coloured operads
but we will henceforth just say operads.
Objects are thought of as
the input and output types of
formal $n$-ary operations
($n \in \mathbb{N}$)
which constitute the morphisms.
If $a, \, b, \, c, \, d, \, e \in \mathrm{Ob}(O)$
are objects of an operad $O$ and
$f \in \mathrm{Hom}(a, \, b; \, c)$,
$g \in \mathrm{Hom}(c, \, d; \, e)$
are morphisms
then the composite
$g \circ \left(f, \, \mathrm{id}_d \right)
 \in \mathrm{Hom}(a, \, b, \, d; \, e)$
corresponds to the following string diagram:
\begin{equation}
	\begin{gathered}
		\includegraphics[width=5cm]{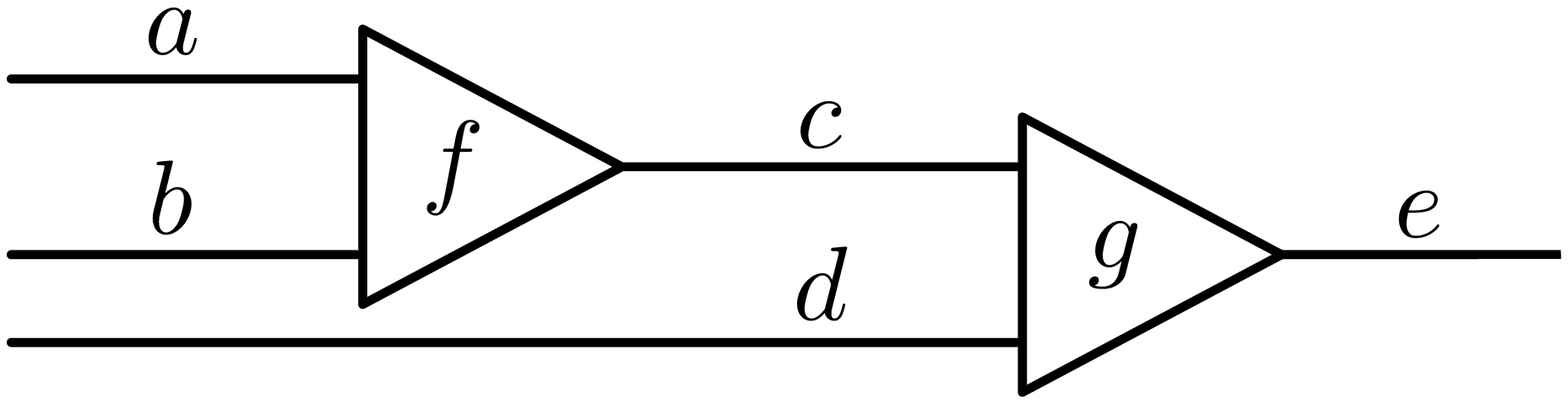}
	\end{gathered}
	\label{eq:operad_morphism}
\end{equation}
The objects of
the operad $\mathrm{Sets}$
are the (small) sets
and its morphisms are all $n$-any functions between them.
%
Functors between operads are defined in essentially the same way.
The formal operations of an operad $O$
can be endowed with semantics
via a functor $F: O \rightarrow \mathrm{Sets}$
called an algebra over $O$.
$F$ sends an object (type)
to the set which contains the elements of that type
and it sends a morphism (formal operation)
to the function which acts on elements of the respective types
in the desired way (i.e.~its meaning or implementation).
The semantics are functorial because
assigning concrete meaning to formal operations
plays together nicely with their composition.

While~\cite{1998Maclane} provides an introduction
to category theory for mathematicians,~%
\cite{2014Spivak} addresses a general scientific audience
and also includes a section on operads and their algebras.
Operads are thoroughly treated in~%
\cite{2004Leinster} and~\cite{2016Yau}.
Operads of wiring diagrams
are extensively covered in~\cite{2018Yau}.

\section{Modelling language}%
\label{sec:language}

Due to their structural properties,
PHS are considered attractive for analysis and control.
EPHS in particular are well suited for multiphysics applications.
Therefore we ask
what is missing to bring PHS to engineering teams?
Although often advertised,
the potential of the framework for such enterprise remains unrealized.
Hoping to make some progress on this front,
we discuss EPHS as a modelling language,
a topic beyond
`equations with geometric structure and thermodynamic interpretation'.


Modelling paradigms are often classified as top-down vs.~bottom-up.
PHS enable a bottom-up approach,
since relatively simple systems
can be interconnected to form more complex ones.
We advocate an operadic approach which
perhaps has more of a top-down flavour:
We argue that modelling of a physical system
starts with its decomposition,
yielding an account
not only of its parts
but also of their interconnection.
Parts can be decomposed recursively
until all low-level parts are sufficiently simple,
making it relatively straightforward
to associate to them their precise meaning.
The structure behind hierarchical decompositions of systems
into interconnected parts
is called the syntax of the modelling language.
The structure which
governs how meaning,
in the form of equations or related data structures,
is associated to decompositions
is called semantics.
Note that we could also conceive of this
as a bottom-up approach:
A syntactic expression then represents
the composition pattern of a system
which can be a part of
a another composition pattern, etc.

\begin{figure}[!htb]
	\centering
	\includegraphics[width=8.4cm]{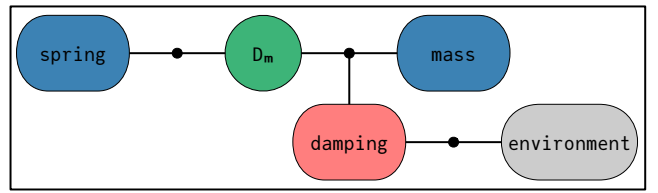}
	\caption{%
		Bond-graph expression
		for defining an EPHS model
		of an isothermal damped harmonic oscillator.
	}%
	\label{fig:oscillator_isothermal}
\end{figure}

It seems rather self-evident that
some sort of bond graphs should be used as
a formal syntax for PHS.
However, as such,
bond graphs are not mathematical objects.
Most researchers probably think of them as directed graphs,
see e.g.~\cite{2020PfeiferCaspartHampelMullerKrebsHohmann}.
In~\cite{2017Coya} a categorical framework is presented
featuring $0$- and $1$-junctions as syntax
and functorial semantics for the thereby expressed junction structure.
%
We propose a categorical framework
with a syntax using only $0$-junctions,
yet arbitrary Dirac structures
can be included via semantics:
A (bond-graph) expression which can be used to define
an isothermal oscillator is shown in
Fig.~\ref{fig:oscillator_isothermal}.

The `outer box' is interpreted as the boundary
of the represented composite system.
The (inner) `boxes'
are drawn with rounded corners or as circles
and represent the system components.
Boxes expose `ports'
which are connected via `bonds' drawn as lines
to `junctions' drawn as black circles.
The junctions
and the box named $\texttt{D}_\texttt{m}$
define the Dirac structure.
The latter connects the potential energy domain on its left
with the kinetic energy domain on its right.
The junction at the bottom right
belongs to the thermal energy domain.

The operad structure allows us
to compose bond-graph expressions.
This can be understood as
substitution of one expression into another.
To define a nonisothermal oscillator,
we need to include the thermal capacity of the damper
as well as a heat transfer law governing its cool-down.
\begin{figure}[!htb]
	\centering
	\includegraphics[width=4.35cm]{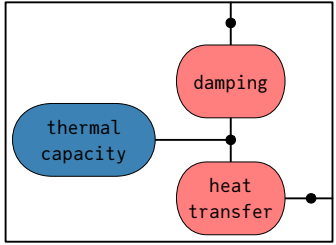}
	\caption{%
		Bond-graph expression for defining an EPHS model
		of a non-isothermal damper.
	}%
	\label{fig:damper_nonisothermal}
\end{figure}
The model refinement can proceed by
substituting the expression
shown in Fig.~\ref{fig:damper_nonisothermal}
into the box labelled $\texttt{damping}$
in the expression shown in Fig.~\ref{fig:oscillator_isothermal}.
The outer box of
the substituted expression
has two `boundary ports'
because it represents an open system.

\begin{figure}[!htb]
	\centering
	\includegraphics[width=7.9cm]{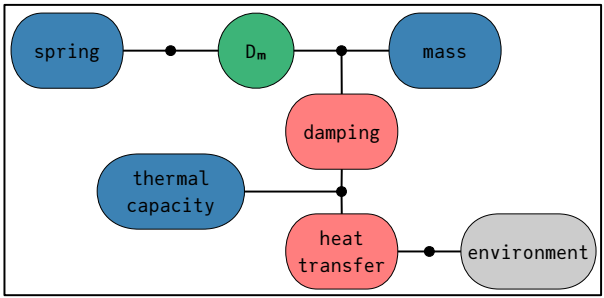}
	\caption{%
		Bond-graph expression for defining an EPHS model
		of a non-isothermal oscillator.
	}%
	\label{fig:oscillator_nonisothermal}
\end{figure}

A box together with its ports
is understood as the interface of
the component that it represents.
Composition
proceeds via identification of interfaces:
The outer box in the substituted expression
is identified with
the box into which it is substituted.
The resulting composite expression is
shown in Fig.~\ref{fig:oscillator_nonisothermal}.
The shared interface disappears
and for each of its ports,
the two corresponding junctions,
one on either side of the interface,
are also identified.

Bonds and their junctions represent the
power-conserving interconnection of components.
At every junction
the flow variables of all incident bonds sum to zero
while the corresponding effort variables are equal.
We fix that
power going into any (outer) box
has a positive sign,
making arrowheads on bonds unnecessary.
This leads to a much simpler formalism
and less visual noise.
Stored power,
dissipated power,
and supplied power
is hence associated with a positive sign.

While junctions incident to
exactly two bonds might seem unnecessary at first,
they are required to formalize a simple and composable syntax.
Further, they are convenient from the perspective of
a graphical user interface (GUI)
for manipulating bond-graph expressions.
Junctions with one incident bond define physically meaningful constraints.

Traditional bond graphs also feature $1$-junctions,
where efforts sum to zero and flows are equal.
While they can be
represented by boxes via semantics,
they seem unnecessary
due to the different roles of
flow and effort variables in the EPHS framework.
Flow variables are related to
rates of extensive quantities
or thermodynamic fluxes,
while effort variables are related to
intensive quantities
or thermodynamic forces.

Boxes are identified via their labels.
Although part of an expression,
we chose to omit labels for (boundary) ports
in their visual representation.
In contrast, junctions do not have labels.

We further omit that
ports and junctions have a type
corresponding to the physical dimensions
of their port variables.
The interconnection of components as captured by expressions
and the operadic composition of expressions
involve automatic type checking.
%
Semantic data defining a low-level component
may further include type annotations
which are checked against its interface.

Although boxes do not have a colour at the level of syntax,
we hint at the associated semantics
by drawing them with a white filling if they represent a nested EPHS.
Similarly, blue for exergy storage,
green for Dirac structure,
red for resistive structure,
and grey for the reference environment.

\begin{figure}[!htb]
	\centering
	\includegraphics[width=7.7cm]{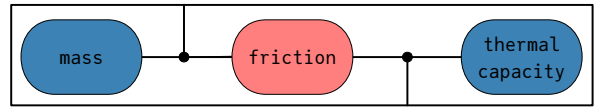}
	\caption{%
		Bond-graph expression for defining an EPHS model
		of a moving piston with friction.
	}%
	\label{fig:piston}
\end{figure}

\begin{figure}[!htb]
	\centering
	\includegraphics[width=8.4cm]{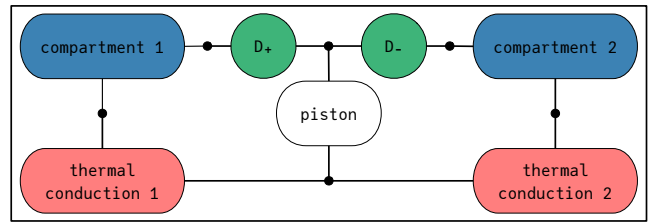}
	\caption{%
		Bond-graph expression for defining an EPHS model
		of a cylinder-piston device.
	}%
	\label{fig:piston_device}
\end{figure}

An expression can be understood as
defining an operation which
takes the meaning of each inner box
and returns the meaning of the outer box.
For instance, the expression
shown in Fig.~\ref{fig:piston}
has two boundary ports.
Hence, if we provide appropriate meaning for its three boxes
we get the meaning of the composite system.
If we then take this as the meaning of
the box labelled $\texttt{piston}$
in Fig.~\ref{fig:piston_device}
and fix an appropriate meaning for the other five boxes,
we obtain the model of the isolated piston-cylinder device
of Example 5.4 in~\cite{2021LohmayerKotyczkaLeyendecker}.
$\texttt{D}_\texttt{+}$
and
$\texttt{D}_\texttt{-}$
interconnect
the kinetic port of the piston
and
the port for pressure-volume work of either compartment,
taking into account the cross-sectional area of the cylinder.
The two Dirac structures differ in sign,
since compartment pressure acts on the piston from different sides.
%
If we would first compose the expressions shown in
Fig.~\ref{fig:piston} and~\ref{fig:piston_device},
i.e.~substitute the former expression
into the box labelled $\texttt{piston}$ in the latter expression,
and then fix the meaning of all boxes
in the composite expression accordingly,
we would obtain the same model
due to functoriality of the semantics.

The environment has a special singleton role.
It is part of every system
in order to define its exergy
and at the same time
it may be represented by any number of boxes,
including zero,
as in the previous example.

\begin{figure}[!htb]
	\centering
	\includegraphics[width=8.4cm]{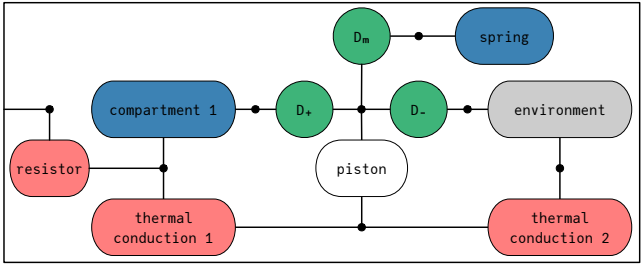}
	\caption{%
		Bond-graph expression for defining an EPHS model
		of a heated cylinder-piston device.
	}%
	\label{fig:piston_device2}
\end{figure}

The operadic approach allows us to hierarchically
define and refine models.
It also allows us to
reuse expressions
either with or without meaning associated to their boxes.
We obtain the model of the heated piston-cylinder device
of Example 5.6 in~\cite{2021LohmayerKotyczkaLeyendecker}
if we reuse parts from the definition of the previous examples
to fix the meaning of the expression
shown in Fig.~\ref{fig:piston_device2}.
Only the resistor
needs to be modelled additionally.

\begin{figure}[!htb]
	\centering
	\includegraphics[width=8.3cm]{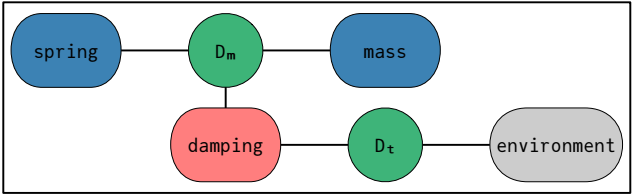}
	\caption{%
		Bond-graph expression for defining an EPHS model
		of an isothermal damped harmonic oscillator
		based on the syntax used in~\cite{2021LohmayerKotyczkaLeyendecker}.
	}%
	\label{fig:oscillator_isothermal_old}
\end{figure}

In~\cite{2021LohmayerKotyczkaLeyendecker}
we used a different syntax,
see Fig.~\ref{fig:oscillator_isothermal_old}.
It is remarkable that
the change required to
make the syntax compatible with the operadic structure
also brings other benefits, namely
a clearer separation of physical domains
and specification of the Dirac structure with fewer boxes.

\subsection{Syntax}%
\label{ssec:syntax}

The syntax is formalized as
the operad of typed undirected wiring diagrams $\mathrm{UWD}$,
see~\cite{2013Spivak} and~\cite{2018Yau}.
It has boxes as objects
and wiring diagrams (i.e.~bond-graph expressions) as morphisms.
%
A box is
a finite set of ports $b$
with a function $\tau_b$ assigning to each port its type:
\begin{equation*}
	\mathrm{Ob}(\mathrm{UWD}) =
	\{
		\left(b, \, \tau_b \right)
		\mid
		b \in \mathrm{Ob}(\mathrm{FinSet}), \,
		\tau_b \colon b \rightarrow \mathrm{Ob}(\mathrm{Set})
	\}
\end{equation*}
For instance,
the box
$b = \{ p_1, p_2 \}$
comes with
a function $\tau_b$ defined by
$\tau_b(p_1) = P_1$,
$\tau_b(p_2) = P_2$.
We think of a port type as
the set of values of flow and effort variables
whose physical dimensions match
the physical domain to which the respective port belongs.
The box can be drawn like so:
\begin{center}
	\includegraphics[width=3.3cm]{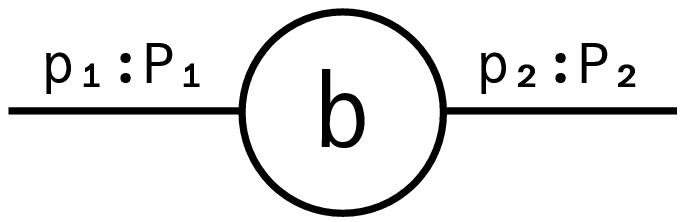}
\end{center}
%
A morphism $f$ is a formal $n$-ary operation
with domain objects or inner boxes $b_i$ ($i = 1, \, \ldots, \, n$)
and codomain object or outer box $b_o$,
i.e. $f \in \mathrm{Hom}(b_1, \, \ldots, \, b_n; \, b_o)$.
For instance,
the morphism $g$
in Eq.~\eqref{eq:operad_morphism}
represents a system
with boundary ports given by the set $e$
and two inner boxes with ports given by the sets $c$ and $d$, respectively.
The composite morphism
$g \circ \left(f, \, \mathrm{id}_d \right)$
represents a system
with boundary ports given by $e$
and three inner boxes with ports given by $a$, $b$, and $d$, respectively.
The data of a $n$-ary morphism,
i.e.~a bond-graph expression with $n$ inner boxes,
is defined by a commutative diagram of the form
\[\begin{tikzcd}
	{\coprod\limits_{i=1}^n b_i} & J & {b_o} \\
	& \textcolor{rgb,255:red,92;green,92;blue,214}{\mathrm{Ob}(\mathrm{Set})}
	\arrow["{\coprod\limits_{i=1}^n \tau_{b_i}}"', color={rgb,255:red,92;green,92;blue,214}, from=1-1, to=2-2]
	\arrow["{\tau_{b_o}}", color={rgb,255:red,92;green,92;blue,214}, from=1-3, to=2-2]
	\arrow["{\mathrm{j}_\mathrm{o}}"', from=1-3, to=1-2]
	\arrow["{\mathrm{j}_\mathrm{i}}", from=1-1, to=1-2]
	\arrow[color={rgb,255:red,92;green,92;blue,214}, from=1-2, to=2-2]
\end{tikzcd}\]
The coproduct $\coprod_i b_i$
is the disjoint union of the ports of all inner boxes.
The functions $\mathrm{j}_\mathrm{i}$
and $\mathrm{j}_\mathrm{o}$
are morphisms in $\mathrm{FinSet}$
and assign ports to junctions.
To express that junctions are not labelled
we can define morphisms as
equivalence classes of diagrams as the above.
However, this seems unnecessary
for software implementation.

In the following we ignore the type information for brevity.
Given a box $b$,
the identity on $b$ is
defined by the diagram
$b \rightarrow b \leftarrow b$
with two identity morphisms in $\mathrm{FinSet}$.

Given
$g \in \mathrm{Hom}(b_1, \, \ldots, \, b_n; \, b_o)$
defined by
\[\begin{tikzcd}
	{\coprod\limits_{i=1}^n b_i} & {J^g} & {b_o}
	\arrow["{\mathrm{j}_\mathrm{i}^g}", from=1-1, to=1-2]
	\arrow["{\mathrm{j}_\mathrm{o}^g}"', from=1-3, to=1-2]
\end{tikzcd}\]
for some
$n \in \mathbb{N}$
and
for each $i \in \{1, \, \ldots, \, n \}$
a morphism
$f_i \in \mathrm{Hom}(b_{i,1}, \, \ldots, \, b_{i,m_i}; \, b_i)$
defined by
\[\begin{tikzcd}
	{\coprod\limits_{k=1}^{m_i} b_{i,k}} & {J^{f_i}} & {b_i}
	\arrow["{\mathrm{j}_\mathrm{i}^{f_i}}", from=1-1, to=1-2]
	\arrow["{\mathrm{j}_\mathrm{o}^{f_i}}"', from=1-3, to=1-2]
\end{tikzcd}\]
for some $m_i \in \mathbb{N}$,
the composite
$g \circ \left( f_i, \, \ldots, \, f_n \right)
 \in \mathrm{Hom}(b_{1,1}, \, \ldots, \, b_{n,m_n}; \, b_o)$
is defined (via pushout) by
\begin{center}
	\begin{tikzcd}
		&&&& {b_o} \\
		&& {\coprod_i b_i} && {J^g} \\
		{\coprod_i \coprod_k b_{i,k}} && {\coprod_i J^{f_i}} && J
		\arrow["{\mathrm{j}_\mathrm{o}^g}", from=1-5, to=2-5]
		\arrow["{\mathrm{j}_\mathrm{i}^g}", from=2-3, to=2-5]
		\arrow["{\coprod_i \mathrm{j}_\mathrm{o}^{f_i}}", from=2-3, to=3-3]
		\arrow["{\coprod_i \mathrm{j}_\mathrm{i}^{f_i}}", from=3-1, to=3-3]
		\arrow["{\mathrm{i}_\mathrm{i}}", from=3-3, to=3-5]
		\arrow["{\mathrm{i}_\mathrm{o}}", from=2-5, to=3-5]
	\end{tikzcd}
\end{center}
with
$\mathrm{i}_\mathrm{i} \circ \left( \coprod_i \mathrm{j}_\mathrm{i}^{f_i} \right)$
and
$\mathrm{i}_\mathrm{o} \circ \mathrm{j}_\mathrm{o}^g$
as the two defining functions.
%
The junctions $J$ of the composite wiring diagram
together with injections $\mathrm{i}_\mathrm{i}$ and $\mathrm{i}_\mathrm{o}$
are determined via the commuting (pushout) square.
The ports of the shared interfaces $\coprod_i b_i$
determine which junctions in $J$ are identified.


Attributed C-Sets
are defined in~\cite{2021PattersonLynchFairbanks}
and
provide a categorical framework
for defining data structures
for graph-like objects with data attributes.
The framework is implemented in the Julia programming language,
see~\cite{2021PattersonFairbanksBaasBrownHalterLibkindLynch}.
Ignoring attributes,
a $C$-Set is a functor from a category $C$ to $\mathrm{FinSet}$.
The combinatorial data of a bond-graph expression
can be expressed as such a functor
with the (free) category $C$ defined by
the black part of
\begin{center}
	\adjustbox{scale=0.9,center}{%
		\begin{tikzcd}
			{\texttt{Port}} && {\texttt{Junction}} && {\texttt{BoundaryPort}} \\
			\\
			{\texttt{Box}} && \textcolor{rgb,255:red,92;green,92;blue,214}{\texttt{PortType}}
			\arrow["{\texttt{ipt}}"', color={rgb,255:red,92;green,92;blue,214}, from=1-1, to=3-3]
			\arrow["{\texttt{jpt}}", color={rgb,255:red,92;green,92;blue,214}, from=1-3, to=3-3]
			\arrow["{\texttt{bpt}}", color={rgb,255:red,92;green,92;blue,214}, from=1-5, to=3-3]
			\arrow["{\texttt{ij}}", from=1-1, to=1-3]
			\arrow["{\texttt{bj}}"', from=1-5, to=1-3]
			\arrow["{\texttt{box}}", from=1-1, to=3-1]
		\end{tikzcd}
	}
\end{center}
Similarly, the violet part is mapped to
the data attributes encoding the port types.

\subsection{Semantics}%
\label{ssec:semantics}

The semantics of the modelling language
is formalized as an algebra over $\mathrm{UWD}$,
i.e.~as a functor to the operad $\mathrm{Sets}$.
On objects,
it sends a box to
the set of EPHS with boundary ports
matching the interface defined by the box.
On $n$-ary morphisms,
it sends a bond-graph expression with $n$ inner boxes to
a function which
takes $n$ EPHS matching the inner boxes
and returns an EPHS matching the outer box.
A bond-graph expression hence defines
a function that composes port-Hamiltonian systems.
Every box in the defining expression
corresponds to a function argument.
Regarding the precise definition of this functor,
the unique role of the environment
deserves special attention.
We defer details of this construction to later.

\section{Conclusion}%
\label{sec:conclusion}

We introduce a formal modelling language
which focuses on expressing physical and compositional structure,
rather than merely systems of equations.
In particular, we show that
a diagrammatic syntax inspired by bond graphs
can be formalized based on
the operad of typed undirected wiring diagrams.
This work
can form the basis of valuable software tools
for researchers working on port-Hamiltonian systems theory.
Eventually,
it could also bring port-Hamiltonian systems
closer to real-world applications.

Future work on the exergetic port-Hamiltonian systems framework
will focus on
presenting a complete definition of the semantics
and on a software implementation of the modelling language.
We also plan to work on
the integration of the modelling language with
distributed-parameter models
and
physics-informed machine learning.


\bibliography{literature}

\end{document}

%% file: abstract.tex
A prevalent theme
throughout science and engineering is
the ongoing paradigm shift
away from isolated systems
towards open and interconnected systems.
Port-Hamiltonian theory developed as
a synthesis of geometric mechanics and network theory.
The possibility to model complex multiphysical systems
via interconnection of simpler components
is often advertised as one of its most attractive features.
The development of
a port-Hamiltonian modelling language however
remains a topic which has not been sufficiently addressed.
We report on recent progress towards
the formalization and implementation of
a modelling language for exergetic port-Hamiltonian systems.
Its diagrammatic syntax inspired by bond graphs
and its functorial semantics
together enable a modular and hierarchical approach to model specification.